\documentclass[conference]{IEEEtran}
\IEEEoverridecommandlockouts
\usepackage{times}
\usepackage{graphicx}
\usepackage{subcaption}
\usepackage{amsmath}
\usepackage{xspace} 
\usepackage{booktabs}
\usepackage{amssymb}
\usepackage{comment}
\usepackage{xcolor}
\usepackage{enumitem} 

\begin{document}
\title{Reservoir Computing-Based Detection for Molecular Communications}
\author{
\IEEEauthorblockN{Abdulkadir Bilge, Eren Akyol, Murat Kuscu}
\IEEEauthorblockA{Nano/Bio/Physical Information and Communications Laboratory (CALICO Lab)\\
Department of Electrical and Electronics Engineering, Ko\c{c} University, Istanbul, 34450, Turkey\\
\texttt{\{abilge20,eakyol22,mkuscu\}@ku.edu.tr}}
}
\maketitle

\begin{abstract}
Diffusion-based Molecular Communication (MC) is inherently challenged by severe inter-symbol interference (ISI). This is significantly amplified in mobile scenarios, where the channel impulse response (CIR) becomes time-varying and stochastic. Obtaining accurate Channel State Information (CSI) for traditional model-based detection is intractable in such dynamic environments. While deep learning (DL) offers adaptability, its complexity is unsuitable for resource-constrained micro/nanodevices. This paper proposes a low-complexity Reservoir Computing (RC) based detector. The RC architecture utilizes a fixed, recurrent non-linear reservoir to project the time-varying received signal into a high-dimensional state space. This effectively transforms the complex temporal detection problem into a simple linear classification task, capturing ISI dynamics without explicit channel modeling or complex retraining. Evaluated in a realistic 3D mobile MC simulation environment (Smoldyn), our RC detector significantly outperforms classical detectors and achieves superior performance compared to complex ML methods (LSTM, CNN, MLP) under severe ISI. Importantly, RC achieves this with significantly fewer trainable parameters (e.g., 300 vs. up to 264k for MLP) and ultra-low latency inference (approx. 1 µs per symbol).

\end{abstract}

\section{Introduction}

Molecular communication (MC) via diffusion is a bio-inspired paradigm where transmitters encode information by modulating the release of signaling molecules. These molecules propagate stochastically toward receivers through Brownian motion. The receiver decodes this information by sensing the concentration of molecules, often via surface receptors, enabling communication in micro- and nanoscale environments where electromagnetic communication is infeasible \cite{kuscu2019transmitter, huang2021signal}. Diffusion-based MC channels, however, exhibit long temporal memory, as molecules from previous transmissions persist and interfere with subsequent symbols, causing inter-symbol interference (ISI) \cite{Jamali2019_Tutorial}. This interference becomes more pronounced as the symbol interval, $T_\mathrm{b}$, decreases, and is further exacerbated in mobile scenarios where the physical channel configuration changes over time.

Traditional ISI mitigation relies on model-based detection, such as Maximum A Posteriori (MAP) sequence detectors \cite{kilinc2014receiver}, requiring accurate channel models \cite{Gomez2025_Survey}. However, in mobile MC, the time-varying channel impulse response (CIR) makes channel state information (CSI) quickly outdated, rendering such approaches ineffective and posing significant challenges to reliable communication \cite{kuscu2019transmitter,araz2023ratio}. Simpler methods, such as fixed-threshold detectors, fail rapidly under heavy ISI due to baseline drift \cite{damrath2016low}. While adaptive thresholding offers slight improvements \cite{damrath2016low}, it cannot compensate for the stochastic nature of ISI in dynamic channels.

Given the intractability of deriving accurate analytical models for dynamic MC environments, researches have explored data-driven alternatives that learn the channel characteristics \cite{huang2021signal,Gomez2025_Survey}. Neural-network-based detectors, including Multilayer Perceptrons (MLPs), have shown improved Bit Error Rates (BER) by implicitly learning ISI correlations \cite{shrivastava2021performance}. Recurrent architectures like Long Short-Term Memory (LSTM) networks and Convolutional Neural Networks (CNNs) can further exploit temporal dependencies, achieving strong performance in both static and mobile environments \cite{gomez2023explainability, Farsad2018}. Nevertheless, these sophisticated ML models require extensive training, high computational cost, and substantial memory, posing severe challenges for energy-limited micro/nanoscale receivers.

In this work, we propose a Reservoir Computing (RC) detector for binary MC that preserves RNN-level temporal modeling at much lower complexity. RC frameworks, exemplified by Echo State Networks (ESNs), use a fixed random recurrent reservoir to map inputs into high-dimensional states while training only a simple linear readout \cite{lukovsevivcius2009reservoir, du2017memristorRC}, providing a nonlinear fading memory that lets the readout capture long-tail ISI and adapt to mobility without full recurrent training. Through extensive simulations in Smoldyn, a particle-based spatial stochastic simulation framework \cite{andrews2017smoldyn}, we demonstrate that RC achieves competitive or superior BER performance compared with conventional and deep-learning-based detectors. This is particularly evident under high-ISI (low-$T_\mathrm{b}$) mobile conditions, while maintaining low model size, training cost, and inference latency, making it a promising approach for practical micro/nanoscale receivers.

\begin{figure*}[!t]
    \centering
    \includegraphics[width=0.9\textwidth, keepaspectratio]{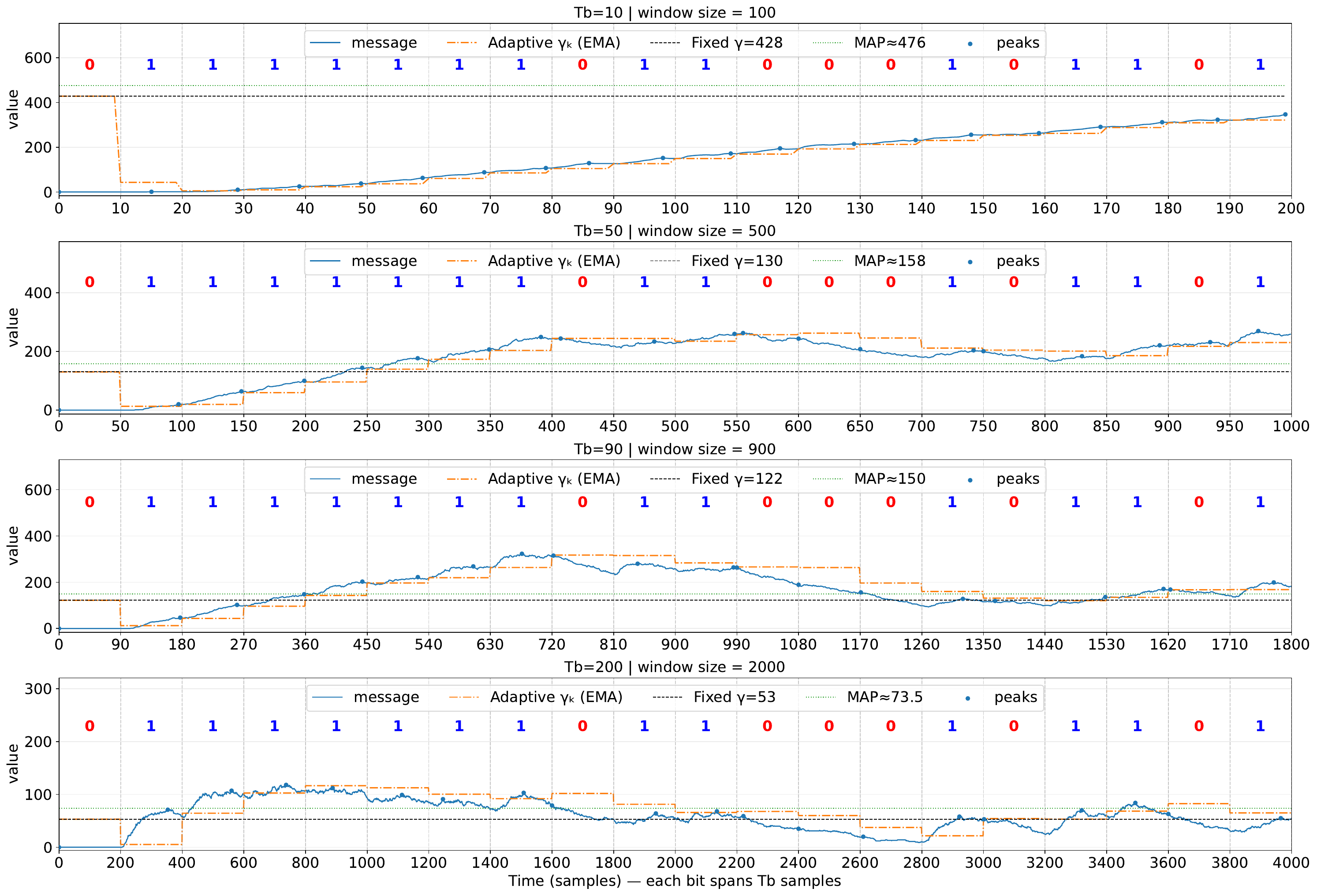}
    \caption{Diffusion-based molecular CIR and accumulated ISI from consecutive symbols. Non-ML detection in diffusion-based MC for different $T_\mathrm{b}$. Blue: received signal; orange: adaptive (EMA) threshold; dotted: fixed threshold.
As $T_\mathrm{b}$ decreases, ISI raises the baseline, and the adaptive detection better tracks it than the fixed threshold.}
    \label{fig:nonml}
\end{figure*}

The main contributions of this paper are:
\begin{itemize}[leftmargin=*]
    \item We propose a novel, lightweight RC-based detector (RC-ISI) specifically designed to handle the time-varying ISI inherent in mobile MC channels.
    \item We provide a comprehensive benchmark of RC against classical (MAP, Adaptive) and modern ML (LSTM, CNN, MLP) detectors in a realistic 3D mobile simulation environment (Smoldyn).
    \item We demonstrate RC's superior robustness under high-ISI conditions and quantify its significant advantages in parameter efficiency (approx. 100x fewer parameters than MLP at $T_\mathrm{b}=100$s) and inference latency, establishing its feasibility for micro/nanoscale implementation.
\end{itemize}

\section{Background}

\subsection{Molecular Communication and the ISI Challenge}

\begin{figure*}[h]
    \centering
    \includegraphics[width=1\textwidth, keepaspectratio]{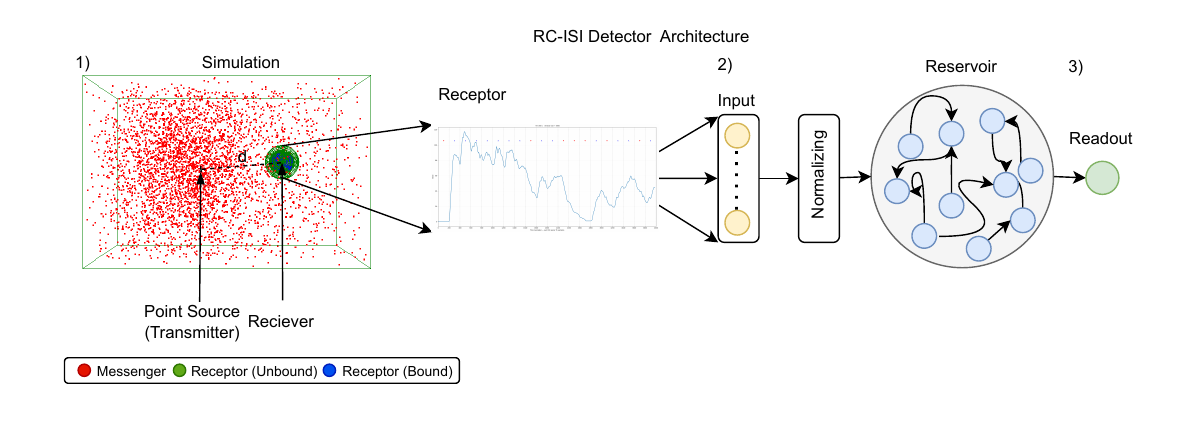}
\caption{Overview of the proposed RC-ISI detection architecture. The mobile MC environment is simulated (1), and the received signal (receptor occupancy) is normalized and fed into the fixed recurrent reservoir (2). The high-dimensional reservoir states are then mapped by the trained readout layer, followed by ROC-based thresholding (3) to produce the final bit decision.}
    \label{fig:smoldyn}
\end{figure*}

In DbMC, information is typically conveyed using Concentration Shift Keying (CSK), such as releasing $N$ molecules for bit $1$ and none for bit $0$ (On-Off Keying). The diffusion process results in a long-tailed CIR \cite{Jamali2019_Tutorial}, meaning molecules from earlier symbols persist in the environment and elevate the baseline concentration for subsequent observations \cite{tepekule2015isi}. When the symbol time $T_\mathrm{b}$ is shortened to increase the data rate, this memory spans many symbols, and ISI becomes the dominant impairment.

Mitigation strategies face significant hurdles. Physical countermeasures, such as introducing strong drift (flow) or deploying degrading enzymes to clear the channel, may not be feasible in many biological environments \cite{Farsad2016_Survey}. Model-based cancellation requires accurate CSI, which is challenging even in static environments, and intractable in mobile ones, and suffers from error propagation. Optimal sequence detectors (e.g., Viterbi/MAP) scale poorly in complexity with the ISI span. While simple peak or fixed-threshold detectors offer low complexity, their performance quickly degrades as $T_\mathrm{b}$ decreases. Consequently, DbMC requires detectors capable of exploiting long temporal dependencies with low complexity, which is the focus of this work. Fig.~\ref{fig:nonml} illustrates the signal accumulation and baseline drift characteristic of high ISI scenarios.

\subsection{Machine Learning–Based Detectors for MC}
Data-driven detectors learn decision rules directly from labeled time traces, avoiding the need for explicit channel estimation. This approach is particularly valuable in mobile or complex environments where the CIR is unknown or time-varying \cite{Gomez2025_Survey}.

Feedforward MLPs and 1-D CNNs can use short windows of recent samples to suppress moderate ISI \cite{shrivastava2021performance}. CNNs are noted for their robustness to channel delays by extracting local temporal features \cite{gomez2023explainability}. Recurrent architectures, such as LSTMs and Gated Recurrent Units (GRUs), maintain an internal state and can, in principle, capture longer memory, making them suitable for sequence detection tasks where ISI is significant \cite{Farsad2018, Gomez2025_Survey}. Bidirectional RNNs (BiRNNs) have also been explored to exploit both past and future samples for decoding the current symbol, aiming to utilize the information contained within the ISI \cite{Farsad2018}. However, these deep models typically require large training sets, careful hyperparameter tuning, and often involve thousands to hundreds of thousands of parameters. This complexity is costly for micro/nanoscale receivers and can lead to poor generalization under rapid channel shifts common in mobile scenarios. By contrast, RC uses a fixed recurrent reservoir as a nonlinear fading memory and trains only a linear readout. This provides temporal modeling capabilities comparable to RNNs but with training complexity closer to shallow models, offering a pragmatic balance between performance and efficiency.

\section{Methodology}

\subsection{System Model and Problem Formulation}
We consider a three-dimensional point-to-point MC system simulated using the Smoldyn framework. A mobile transmitter (TX) emits binary symbols $\{b_k\} \in \{0, 1\}$ over intervals of duration $T_\mathrm{b}$. For $b_k=1$, the TX instantaneously releases $N$ messenger molecules; for $b_k=0$, no molecules are released.

\textbf{Mobility Model:} Both the TX and the receiver (RX) are mobile within a bounded cubic domain. They move according to Brownian motion, characterized by their respective diffusion coefficients, $D_\mathrm{T}$, $D_\mathrm{R}$. Boundaries are reflective.

\textbf{Channel and Reception:} A spherical RX with reactive surface receptors is located within the volume. Messenger molecules propagate via Brownian diffusion (coefficient $D_\mathrm{M}$). Molecules may bind to RX receptors with a forward rate $k_f$ and unbind with a backward rate $k_b$. The received signal $r(t)$ is defined as the time series of the number of bound receptors at the RX surface. Due to the stochastic nature of diffusion and the mobility of both TX and RX, the channel is time-varying. The received signal at time $t$ can be modeled as:
\[
r(t) = \sum_{k} s_k(t) * h(t, \tau_k) + n(t),
\]
where $s_k(t)$ is the transmitted signal for symbol $k$, $n(t)$ represents noise (including diffusion and ligand-receptor binding noise), and $h(t, \tau_k)$ is the time-varying CIR from the TX at emission time $\tau_k$ to the RX at time $t$. The mobility causes $h(t, \tau_k)$ to change continuously, making CSI acquisition intractable. Moreover, the long-tailed nature of $h(t, \tau_k)$ results in strong ISI.

\begin{table}[htbp]
\centering
\caption{Simulation Parameters}
\label{tab:sim_params_detailed}
\begin{tabular}{@{}ll@{}}
\toprule
\textbf{Parameter} & \textbf{Value} \\
\midrule
Molecules per bit-1 pulse ($N$) & $2000$ \\
Molecule Diff. Coeff. ($D_\mathrm{M}$) & $1.01 \times 10^{-9}\, \mathrm{m^2/s}$ \\
TX Diff. Coeff. ($D_\mathrm{T}$) & $4.74 \times 10^{-14}\, \mathrm{m^2/s}$ \\
RX Diff. Coeff. ($D_\mathrm{R}$) & $2.31 \times 10^{-12}\, \mathrm{m^2/s}$ \\
RX Radius ($r_0$) & $5\, \mu\mathrm{m}$ \\
Forward binding rate ($k_f$) & $12.5 \times 10^{-15}\, \mathrm{\frac{m^3}{molecule \cdot s}}$ \\
Backward binding rate ($k_b$) & $1000\, \mathrm{s^{-1}}$ \\
Simulation time step ($T_s$) & $0.1\, \mathrm{s}$ \\
\bottomrule
\end{tabular}
\end{table}

\textbf{Problem Formulation:} For symbol $k$, we extract an input feature $u_k$. In this work, $u_k$ is defined as the sampled receptor occupancy at the end of the symbol interval, i.e., $u_k = r(k T_\mathrm{b})$. The task of the detector is a binary classification with memory: estimate the transmitted bit $\hat{b}_k$ given the sequence of current and past observations $\{u_1, u_2, \ldots, u_k\}$. Our goal is to design a detector that maximizes detection accuracy, particularly focusing on scenarios where ISI is strong and the channel is dynamic. Simulation parameters used in this study are summarized in Table~\ref{tab:sim_params_detailed}.

\subsection{RC-Based Detector for MC}

The proposed detector utilizes an Echo State Network (ESN), a specific type of RC architecture, composed of a fixed recurrent reservoir and a trainable linear readout (see Fig.~\ref{fig:smoldyn}). We investigate two detection strategies based on this architecture: standard RC and an enhanced RC-ISI.

 \textbf{Reservoir Layer:}
The reservoir is an untrained recurrent network of $N_r$ nonlinear neurons with a state vector $\mathbf{x}_k \in \mathbb{R}^{N_r}$. We employ a Leaky-Integrator ESN (LI-ESN), which is well-suited for tasks with varying time scales. Following \cite{gauthier2021nextgen}, the state evolves according to:
\[
\mathbf{x}_k = (1-\alpha)\mathbf{x}_{k-1} + \alpha \tanh(W_{\text{res}}\mathbf{x}_{k-1} + W_{\text{in}}u_k),
\]
where $u_k$ is the normalized input feature at step $k$, $\alpha \in (0, 1]$ is the leaky rate, $W_{\text{in}}$ is the input weight matrix, and $W_{\text{res}}$ is the recurrent weight matrix.

$W_{\text{in}}$ and $W_{\text{res}}$ are randomly initialized. $W_{\text{res}}$ is scaled such that its spectral radius $\rho(W_{\text{res}}) < 1$ to ensure the echo-state property, guaranteeing that the reservoir state is a function only of the input history. These fixed recurrent connections form a nonlinear fading memory that encodes the complex ISI and channel dynamics into a high-dimensional trajectory $\mathbf{x}_k$.

\textbf{Feature Normalization:}
Before entering the reservoir, the input features $u_k$ are normalized (z-score normalization) based on training data statistics. This removes amplitude bias and highlights relative fluctuations, which is critical under ISI where the signal baseline can drift unpredictably.

\textbf{Readout Training and Detection Strategies:}
The readout layer maps the reservoir state $\mathbf{x}_k$ to the output. The output weights $\mathbf{w} \in \mathbb{R}^{N_r+1}$ (including bias) are the only parameters trained. We use Ridge Regression to map the reservoir state $\mathbf{x}_k$ to the target bit $b_k$. This is a convex optimization problem with a closed-form solution, avoiding iterative backpropagation \cite{lukovsevivcius2009reservoir}. We compare two detection strategies:

\textbf{\textit{1) Standard RC Detection:}} The decision is made using a fixed threshold of 0.5 on the regression output: $\hat{b}_k = 1$ if $\mathbf{w}^\top\mathbf{x}_k > 0.5$.

\textbf{\textit{2) RC-ISI Detection:}} To enhance robustness against varying ISI levels and channel dynamics, we optimize the decision threshold. We analyze the Receiver Operating Characteristic (ROC) curve on validation data and select the optimal threshold $\eta_{opt}$ that minimizes the BER. The decision is $\hat{b}_k = 1$ if $\mathbf{w}^\top\mathbf{x}_k > \eta_{opt}$.

\begin{figure}[t]
    \centering
    \includegraphics[width=0.7\columnwidth]{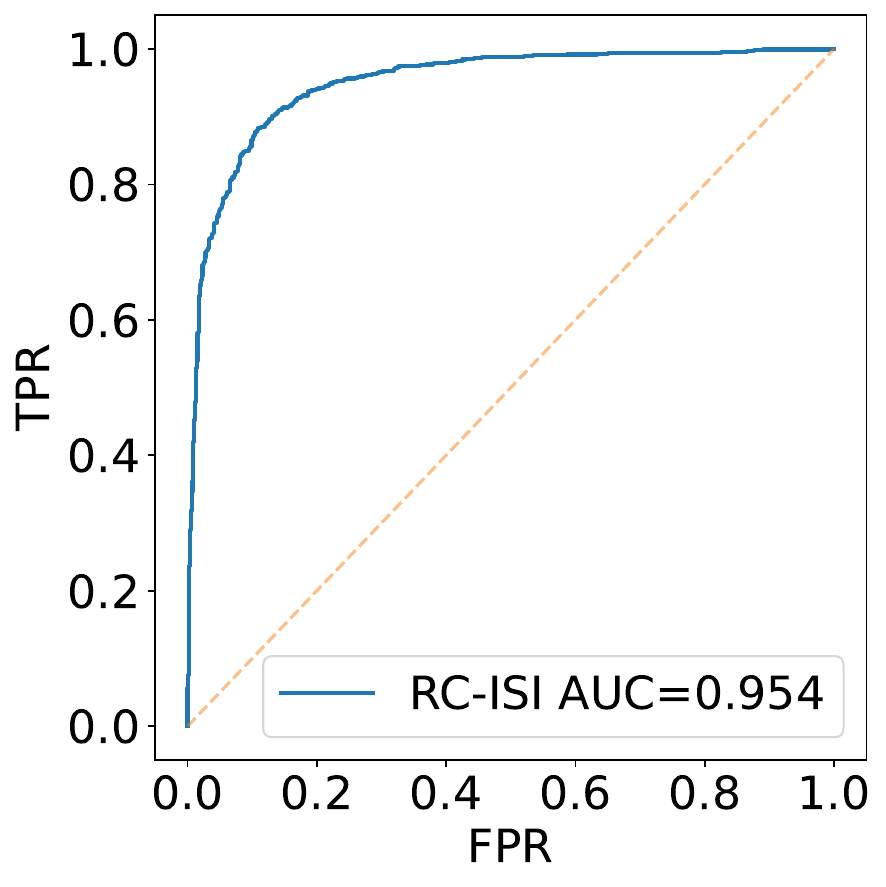}
    \caption{ROC curve for RC-ISI at $T_\mathrm{b}{=}100$s. The high AUC (0.954) indicates strong classification performance and separability between bit $1$ and bit $0$.}
    \label{fig:roc_curve}
\end{figure}

Fig.~\ref{fig:roc_curve} illustrates the ROC curve for RC-ISI at $T_\mathrm{b}=100$s, demonstrating strong classification confidence (AUC 0.954).

\textbf{Hyperparameters:}
We empirically fine-tuned the hyperparameters for RC-ISI. The settings used are: Reservoir size $N_r=400$, spectral radius $\rho=0.7$, leaky rate $\alpha=0.3$, input scaling $s_{\text{in}}=1.0$, and washout length (initial steps discarded during training) $T_{\text{wash}}=300$. These parameters ensure sufficient memory capacity and stability for mobile high-ISI scenarios.

\subsection{Baseline Detectors for Comparison}
To benchmark the RC-based detectors, we evaluate several representative baselines from the MC literature \cite{kuscu2019transmitter, huang2021signal, Gomez2025_Survey}.

\textbf{Fixed Threshold Detector (Peak-Fixed):}
A conventional single-sample detector that compares the received signal $u_k$ (sampled at the peak time, if known, or end of the interval) to a fixed threshold $\eta$, tuned for minimum BER on training data ignoring the channel memory.

\textbf{Adaptive Detector (Adaptive-EMA):}
Following the low-complexity scheme in~\cite{damrath2016low}, the threshold dynamically adjusts to residual ISI via an exponential moving average (EMA) of past observations:
\[
\tilde{I}_k = \beta \tilde{I}_{k-1} + (1-\beta)u_{k-1},
\]
where $\beta$ is the smoothing factor and the decision is $\hat{b}_k=1$ if $u_k > \eta+\tilde{I}_k$. This improves robustness to baseline drift.

\textbf{MAP Sequence Detector:}
A classical model-based detector implemented via the Viterbi algorithm. Since instantaneous CSI is unavailable in this mobile environment, we implemented a mismatched MAP detector assuming a static channel with an average CIR estimated from the training data. While theoretically optimal with perfect CSI, its performance is expected to degrade significantly when channel assumptions are violated due to mobility \cite{kilinc2014receiver}.

\textbf{Feedforward-MLP:}
A data-driven detector using a one-hidden-layer MLP with input window $[u_k,\dots,u_{k-W+1}]$, ReLU activations, and a sigmoid output. Via backpropagation, it learns nonlinear ISI compensation.

\textit{Architecture Details:} 
We employ a feed-forward neural network (MLP) with an input window of size $W = 10T_\mathrm{b}$ (single-channel), 
two hidden layers $H = [128,\,64]$ using ReLU activation, and a single logistic output neuron. 
The total number of trainable parameters is given by $
N_p = W \times 128 + 128 \times 64 + 64 \times 1 + (128 + 64 + 1) = 128W + 8449$, which scales linearly with $T_b$. 
As $T_b$ increases from $10$ to $200$, the input window $W$ expands from $100$ to $2000$, 
resulting in an increase in parameter count from approximately $21\text{k}$ to $264\text{k}$. 
The window size $W$ is scaled proportionally to $T_b$ to capture the relevant ISI span, 
while the hidden layer sizes are kept constant to isolate the effect of input context length.

\textbf{Artificial Neural Network (ANN):} A data-driven detector based on a shallow ANN with an input window $[u_k,\dots,u_{k-W+1}]$, ReLU activations, and a sigmoid output. 
Trained via backpropagation, it performs nonlinear mapping for ISI mitigation 
while maintaining low complexity compared to deeper MLP architectures.

\textit{ANN Architecture: }
We use a shallow feed-forward network with input window $W$ (single-channel), flattened and passed to two hidden layers of sizes $d_1$ and $d_2$ with ReLU, followed by a sigmoid output. In our setup we fix $d_1=32$ and $d_2=16$; $W$ scales proportionally with $T_\mathrm{b}$ to match the ISI span (same $W$ policy as the MLP). The trainable parameter count is
$N_p \;=\; W d_1 \;+\; d_1 \;+\; d_1 d_2 \;+\; d_2 \;+\; d_2 \;+\; 1$,
which with $d_1{=}32,d_2{=}16$ simplifies to $N_p = 32W + 577$. 
By comparison, the MLP in Table~II uses larger hidden sizes $[128,64]$ and thus scales as $N_p = 128W + 8449$, yielding substantially more parameters for the same $W$.

\textbf{Convolutional Neural Network (CNN):}
A lightweight 1-D CNN (two layers, small kernels) is examined~\cite{gomez2023explainability}. CNNs extract local temporal features and are relatively robust to time shifts. The architecture used has $465$ parameters across all $T_\mathrm{b}$.

\textbf{Long Short-Term Memory (LSTM):}
An LSTM network ($H_l{=}16$–32 units) is examined. LSTMs model long-term dependencies but require expensive backpropagation-through-time. The architecture used has $5$k parameters across all $T_\mathrm{b}$.

\begin{table*}[!t]
\vspace{1ex}
\caption{Benchmark Summary: Accuracy, Parameter Count, and Inference Time (µs) Across Symbol Intervals ($T_\mathrm{b}$).}
\label{tab:bench_full}
\centering
\setlength{\tabcolsep}{3pt}
\renewcommand{\arraystretch}{1.07}
\begin{tabular}{lrrrrrrrrrrrrrrrrrrrrr}
\toprule
 & \multicolumn{7}{c}{\textbf{Accuracy(\%)}} & \multicolumn{7}{c}{\textbf{Parameter Count}} & \multicolumn{7}{c}{\textbf{Inference Time (µs)}} \\
\cmidrule(lr){2-8}\cmidrule(lr){9-15}\cmidrule(lr){16-22}
Model\hspace{1 cm}    $T_\mathrm{b}$(s)= & 10 & 30 & 50 & 70 & 90 & 100 & 200 & 10 & 30 & 50 & 70 & 90 & 100 & 200 & 10 & 30 & 50 & 70 & 90 & 100 & 200 \\
\midrule
Peak-Fixed & 50 & 49 & 53 & 56 & 58 & 56 & 65 & - & - & - & - & - & - & - & [N/A] & [N/A] & [N/A] & [N/A] & [N/A] & [N/A] & [N/A] \\
Adaptive-EMA & 52 & 56 & 65 & 70 & 83 & 71 & 83 & - & - & - & - & - & - & - & [N/A] & [N/A] & [N/A] & [N/A] & [N/A] & [N/A] & [N/A] \\
MAP (Mismatched) & 50 & 49 & 54 & 57 & 57 & 56 & 59 & - & - & - & - & - & - & - & [N/A] & [N/A] & [N/A] & [N/A] & [N/A] & [N/A] & [N/A] \\
\midrule
CNN & 52 & 48 & 52 & 58 & 52 & 56 & 63 & 465 & 465 & 465 & 465 & 465 & 465 & 465 & 103.9 & 105.7 & 117.1 & 217.7 & 167.9 & 117.6 & 203.7 \\
LSTM & 51 & 52 & 55 & 69 & 69 & 61 & 73 & 5k & 5k & 5k & 5k & 5k & 5k & 5k & 306.4 & 501.6 & 754.6 & 1300 & 1300 & 1300 & 2800 \\
ANN & 52 & 65 & 77 & 90 & 89 & 87 & 97 & 5k & 11k & 18k & 24k & 30k & 34k & 66k & 92.4 & 90.8 & 98.4 & 178.5 & 130.3 & 89.62 & 170.2 \\
Feedforward-MLP & 52 & 54 & 77 & 85 & 90 & 86 & 95 & 21k & 47k & 72k & 98k & 124k & 136k & 264k & 0.68 & 0.84 & 0.96 & 39.95 & 1.41 & 1.40 & 13.93 \\
\midrule
\textbf{RC} & 53 & 64 & 81 & 87 & 86 & 83 & 90 & 301 & 301 & 301 & 301 & 301 & 301 & 301 & 0.45 & 0.97 & 0.49 & 3.92 & 0.55 & 0.47 & 0.85 \\
\textbf{RC-ISI} & 56 & 65 & 80 & 91 & 89 & 87 & 98 & 401 & 401 & 401 & 401 & 401 & 401 & 401 & 0.99 & 0.86 & 1.01 & 6.82 & 1.01 & 0.97 & 1.20 \\
\bottomrule
\end{tabular}
\end{table*}

\section{Results and Discussion}

We evaluated the proposed RC detectors across ISI levels by varying the symbol duration $T_\mathrm{b} \in \{10, 30, 50, 70, 90, 100, 200\}$ in the 3D mobile MC simulation environment described in Section~III-A. Each configuration was trained and tested using identical bit sequences generated within the Smoldyn simulation, and all detectors used identical input features (sampled receptor occupancy) for a fair comparison. We report accuracy, number of trainable parameters, and inference time.

\begin{figure}[t]
  \centering
  \includegraphics[width=0.9\linewidth]{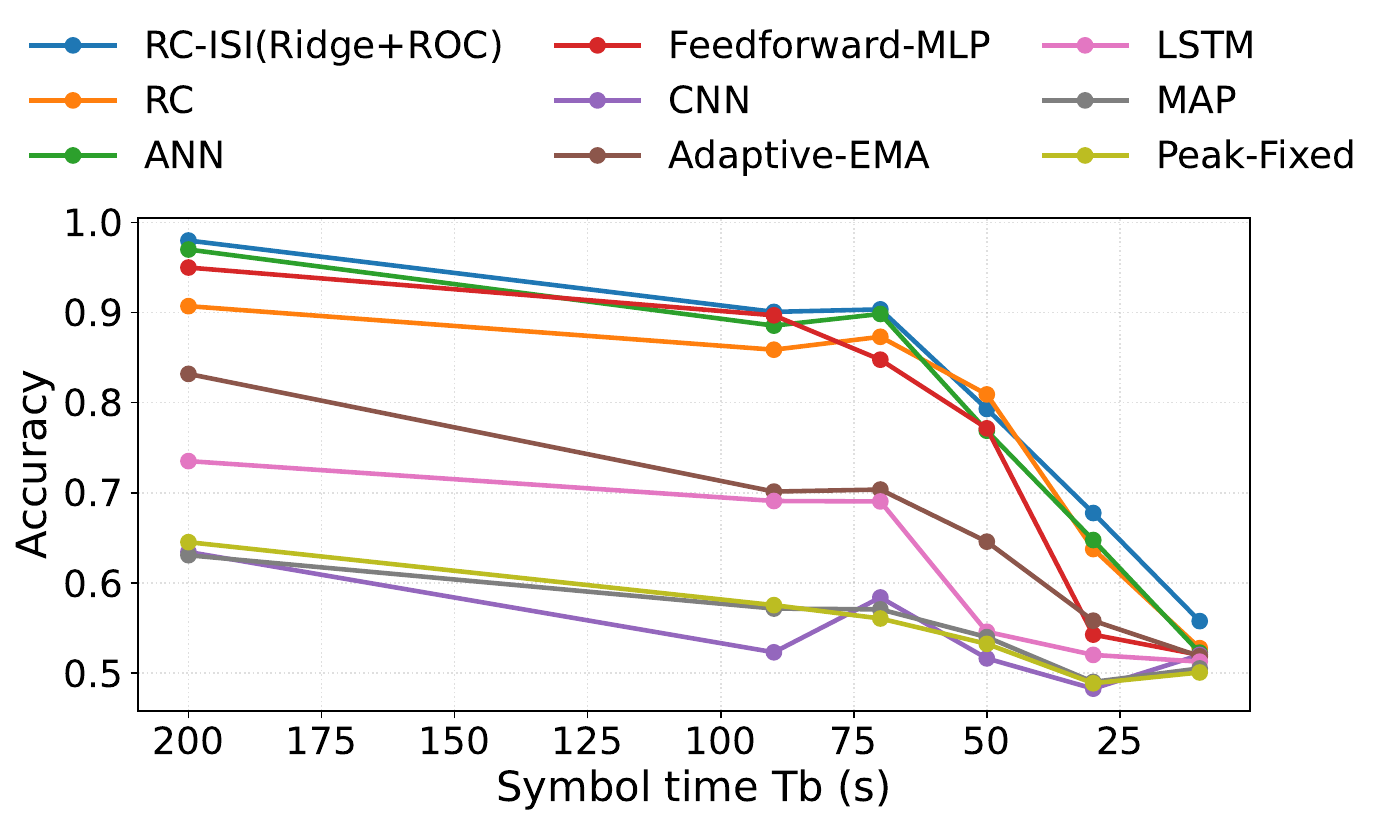}
  \caption{Accuracy comparison versus symbol time ($T_\mathrm{b}$) for all evaluated detectors. RC-ISI maintains the highest robustness as $T_\mathrm{b}$ decreases (ISI increases), significantly outperforming classical methods and complex ML models under severe ISI.}
  \label{fig:accuracy_tb}
\end{figure}

Across ISI levels, Table~\ref{tab:bench_full} summarizes detection accuracy and Fig.~\ref{fig:accuracy_tb} visualizes the trends. As $T_\mathrm{b}$ decreases and ISI strengthens, all methods degrade, yet the RC variants maintain a clear advantage. At the most challenging condition ($T_\mathrm{b}{=}10$\,s), RC-ISI attains \textbf{$56$\%} accuracy, while other ML models and threshold-based detectors approach random performance ($49$–$52$\%). The mismatched MAP detector sits near $50$\%, emphasizing the fragility of classical model-based approaches when perfect or static CSI is unavailable in mobile environments. At moderate ISI ($T_\mathrm{b}{=}30$\,s), RC-ISI reaches \textbf{$65$\%}, outperforming MLP ($54$\%), CNN ($48$\%), and LSTM ($52$\%). As ISI relaxes ($T_\mathrm{b}{=}70$–$90$\,s), RC-ISI achieves high accuracy (around $90$\%), comparable to the best-performing feedforward networks (ANN and MLP). At low ISI ($T_\mathrm{b}{=}100$–$200$\,s), performance converges across several learned models; for example, at $T_\mathrm{b}{=}100$\,s we observe RC-ISI ($87$\%), ANN ($87$\%), and MLP ($86$\%). Overall, the RC detectors remain well above chance even under the strongest ISI, demonstrating robust temporal generalization, with RC-ISI’s threshold optimization providing additional gains.

These accuracy results reveal why RC is effective in time-varying channels. The reservoir acts as a nonlinear dynamic memory that projects stochastic, time-varying inputs into a high-dimensional state where symbols become more linearly separable, allowing a simple linear readout to draw decision boundaries that remain robust to mobility-induced CIR variations. This stands in contrast to threshold-based detectors that treat ISI as noise~\cite{damrath2016low} and to MAP detectors that rely on accurate CSI. The near-random performance of the mismatched MAP detector (Table~\ref{tab:bench_full}) makes this distinction explicit.

\begin{figure}[t]
  \centering
  \includegraphics[width=0.9\linewidth]{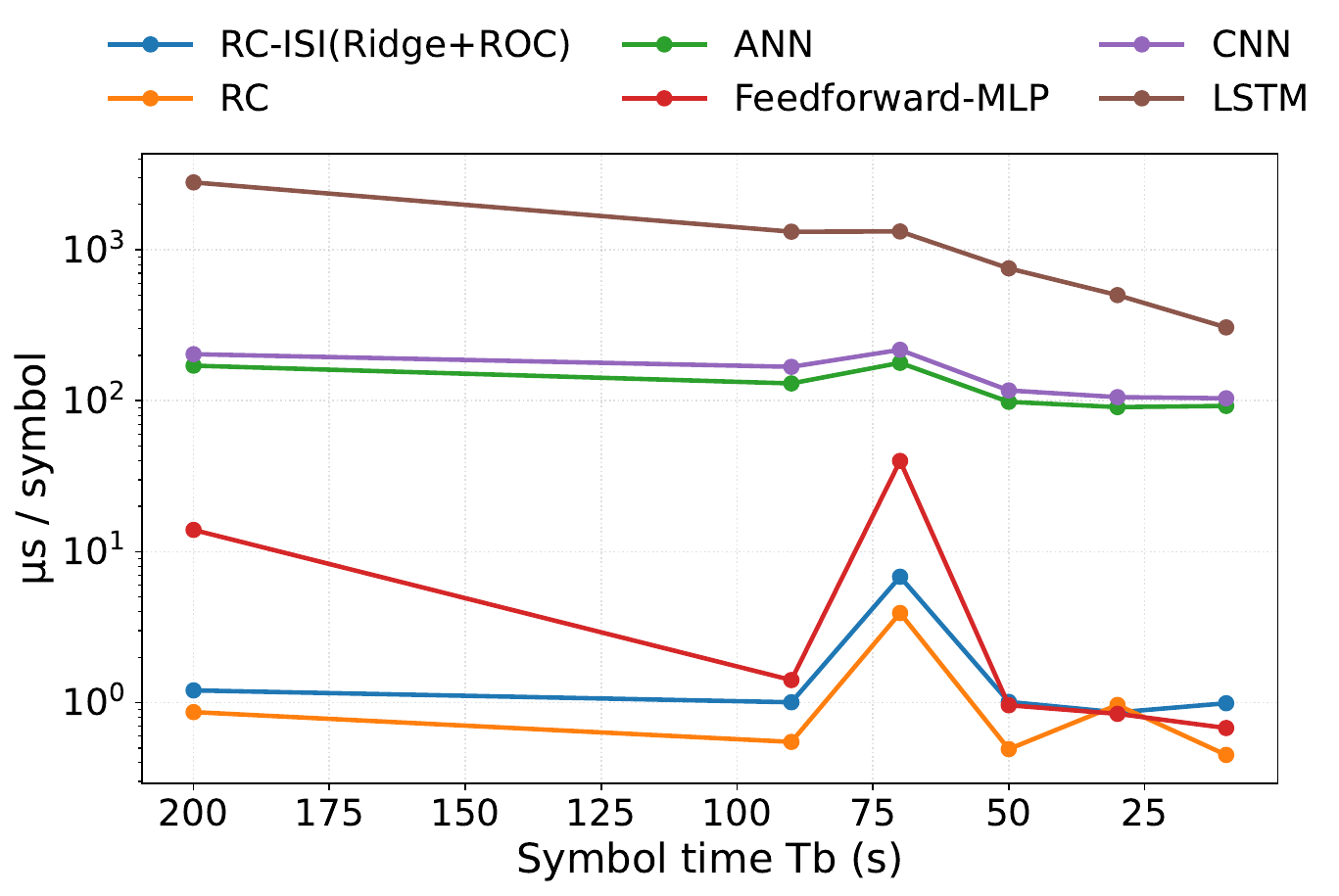}
  \caption{Inference latency ($\mu$s, log scale) of detectors across different symbol durations. RC and RC-ISI achieve the lowest latency among ML methods, comparable to simple MLP implementations and significantly faster than LSTM/CNN.}
  \label{fig:latency_tb}
\end{figure}

Computational complexity and latency further differentiate the methods. RC-ISI employs only 401 trainable parameters (the readout layer), whereas the MLP scales between $21$k and $264$k parameters depending on $T_\mathrm{b}$, and the ANN requires up to $66$k parameters. LSTM uses about $5$k parameters yet incurs significantly higher training complexity due to backpropagation-through-time. As shown in Fig.~\ref{fig:latency_tb}, RC and RC-ISI deliver ultra-low inference latency: RC-ISI processes a symbol in roughly $1\,\mu$s (from $0.86\,\mu$s to $6.82\,\mu$s), vastly faster than LSTM ($306\,\mu$s to $2800\,\mu$s) and CNN ($103\,\mu$s to $217\,\mu$s). This efficiency arises from RC’s structural simplicity such that each decision requires only a fixed-cost reservoir update and a single linear readout step. Compared to fully trained recurrent networks such as LSTMs, RC attains superior accuracy in high-ISI regimes with far lower training cost, avoiding vanishing/exploding gradients and heavy data demands~\cite{Gomez2025_Survey}. Feedforward models (MLP, ANN) had to scale substantially (up to $264$k parameters) to span varying ISI levels, whereas RC maintained an efficient performance with 401-parameter readout. The resulting low memory footprint, fast training, and low inference time underscore hardware feasibility for resource-constrained micro/nanodevices and even suggest compatibility with emerging wetware realizations~\cite{baltussen2024chemical,perera2025wet}. In short, the method combines high-ISI robustness and mobility adaptation (e.g., $>55$\% accuracy at $T_\mathrm{b}{=}10$\,s), ultra-low inference latency (approximately $1\,\mu$s per symbol), and extreme parameter efficiency (401 trainable weights), achieving a balance among accuracy, complexity, and latency.

\section{Conclusion}

We proposed a RC-based detector for mobile diffusion-based MC under severe ISI. Simulations demonstrated that the RC-ISI detector sustains superior robustness and higher relative accuracy compared to classical detectors (MAP, Adaptive) and complex ML methods (LSTM, CNN, MLP) when ISI is severe. Importantly, RC achieves this performance with a fraction of the complexity, utilizing only 401 trainable parameters and offering microsecond-level inference latency. Its closed-form training and small footprint make it highly suitable for real-time operation on resource-constrained micro/nanodevices. Although RC demonstrates superior robustness, its accuracy at $T_\mathrm{b}=10$ (56\%, BER $0.44$) is insufficient for reliability, necessitating strong error-correction coding. Therefore, future work should optimize reservoir topology (e.g., channel-informed initialization) and enable adaptive online readout retraining for rapid channel variations.

\section*{Acknowledgment}
We thank Saad Yousuf for his support in Smoldyn simulations. This work was supported by The Scientific and Technological Research Council of Turkey (TUBITAK) under Grants \#123E516 and \#123C592.

\bibliographystyle{IEEEtran}
\bibliography{references}
\end{document}